\documentclass[11pt,a4paper]{article}
\usepackage{amsmath,amsthm}
\usepackage{amssymb}

\usepackage{graphicx}

\newtheorem{theorem}{Theorem}[section]
\newtheorem{proposition}[theorem]{Proposition}

\newtheorem{corollary}[theorem]{Corollary}

\theoremstyle{definition}
\newtheorem{definition}[theorem]{Definition}

\def\pd#1#2{\frac{\partial#1}{\partial#2}}
\def\core#1{{{#1}_*}}

\begin{document}


\title{Revisiting Lie integrability by quadratures \\ from a
geometric perspective}
\author{Jos\'e F. Cari\~nena\footnote{email: jfc@unizar.es} \\
\textit{Departamento de F\'{\i}sica Te\'orica and IUMA}\\
\textit{Facultad de Ciencias,
 Universidad de Zaragoza} \\
\\
Fernando  Falceto\footnote{email: falceto@unizar.es}\\
\textit{Departamento de F\'{\i}sica Te\'orica and BIFI}\\
\textit{Facultad de Ciencias,
 Universidad de Zaragoza}
 \\
\\
Janusz Grabowski\footnote{email: jagrab@impan.pl}\\
\textit{Institute of Mathematics, Polish Academy of Sciences} \\
\\ Manuel F. Ra\~nada\footnote{email: mfran@unizar.es}\\
\textit{Departamento de F\'{\i}sica Te\'orica and IUMA}\\
\textit{Facultad de Ciencias,
 Universidad de Zaragoza} 
}
\date{}

\maketitle

\begin{abstract}
After a short review of  the classical Lie theorem,  a  finite dimensional
Lie algebra of vector fields is considered and  the most general conditions under which
the integral curves of one of the fields can be obtained by quadratures in a prescribed
way will be discussed, determining also the number of quadratures needed to integrate the system.
  The theory will be illustrated with examples and
 an extension of the theorem where the Lie algebras are replaced by some distributions will also be presented.
\end{abstract}

\section{Introduction: the meaning of  Integrability}

Integrability is a topic that has been receiving quite a lot of attention because such not clearly defined notion appears in many branches of science, and in particular in physics.
The exact meaning of integrability is only well defined in each specific field and each one of the many possibilities of defining in a precise way the concept of integrability has
 a theoretic interest. Loosely speaking integrability refers to the possibility of finding the solutions of a given differential equation (or a system of differential equations),
  but one may also look for solutions of certain types, for instance, polynomial or rational ones, or expressible in terms of elementary functions. The existence
  of additional geometric structures allows us to introduce other concepts of integrability, and so the notion of integrability   is often identified as complete integrability
   or Liouville integrability \cite{arnold}, but we can also consider generalised Liouville
integrability or even non-Hamiltonian integrability \cite{MF78}. For a recent description of other related integrability approaches see e.g. \cite{olver,MCSL14}.

Once a definition of integrability is accepted, systems are classified into integrable and nonintegrable systems. Groups of equivalence
 transformations allow to do a finer classification, all systems in the same orbits having the same integrability properties. Therefore if some
 integrable cases have been previously selected we will have a related family of integrable cases. So, even if the generic Riccati equation is
  not integrable by quadratures, all Riccati  equations related to inhomogeneous linear differential equations are integrable by quadratures too, and
   this provides us integrability conditions for Riccati equations \cite{CdLa,CdLR,CR99}.

The knowledge of particular solutions can also be useful for transforming the original system  in simpler ones, and the prototypes
of this situation are the so called Lie systems admitting a superposition rule for expressing their general solutions in terms of a generic
set of a finite number of solutions \cite{CGM00,CGM01,CGM07,CGR01,CIMM15,CdLb,CR99}.
 This is a report
of a recent collaboration  Prof. Grabowski with members of the Department of Theoretical Physics of Zaragoza University \cite{CFGR15} on a different concept
of integrability, the most classical Lie concept of integrability by quadratures, i.e. all of its solutions can be found  by algebraic operations
 (including inversion of functions) and computation of integrals of functions of one variable (called quadratures).

 Our approach does not resort
  to the existence of additional compatible structures, but simply uses modern tools of algebra and geometry. In order to avoid dependence of a particular  choice of coordinates we should consider the
   problem from a geometric perspective, replacing the systems of differential equations by
   vector fields, a global concept, in such a way that the integral curves of such vector fields are the solutions of a system of differential equations in a coordinate system.
  The two main tools to be used are
  finite-dimensional Lie algebras of vector fields, in particular solvable Lie algebras (see e.g. \cite{arnoldkozlov}) or nilpotent  Lie algebras \cite{MK88,G90},  and distributions spanned by vector fields. The aim is to extend
  Lie classical results of integrability \cite{arnoldkozlov}.

The paper is organised as follows: the fundamental notions on Lie integrabilty and their relations with the standard Arnold-Liouville
integrability are recalled in
Section 2 and some  concepts of cohomology needed to analyse the existence of solutions for a system of first order differential equations are recalled in
 Section 3. The approach to integrability recently proposed in \cite{CFGR15} is sketched in
 section  4
 and some interesting algebraic properties are studied
in Section 5.  The approach is illustrated  in section 6 with the analysis, without any recourse to the symplectic structure, of a recent example of a Holt-related potential that is not separable
 but is superintegrable with high order first integrals, while the last sections are  devoted to extending the previous results to the more general situation in which, instead of having a Lie algebra, $L$,  of vector fields, we have a vector space $V$ such that its elements do not close a finite dimensional real Lie algebra, but rather generate a  general integrable distribution of vector fields.

\section{Integrability by quadratures}

Given an autonomous  system of first-order differential equations,
\begin{equation}
\dot x^i=  f^i(x^1,\ldots,x^N)\ ,\qquad  i=1,\ldots,N, \label{autsyst}
\end{equation}
we can consider changes of coordinates and then the system (\ref{autsyst}) becomes a
new one. This suggest that (\ref{autsyst})
 can be  geometrically interpreted in terms of a vector field $\Gamma$ in a
$N$-dimensional manifold $M$
whose local  expression in the given coordinates is
$$
\Gamma=f^i(x^1,\ldots,x^N)\pd{}{x^i}\ .\label{leavf}
$$
The  integral curves of $\Gamma$ are  the  solutions of the given system,
and then integrate the system means
to  determine the general solution of the system.
More specifically,  integrability by quadratures means that you can determine the solutions (i.e. the flow of $\Gamma$) by means of a
 finite number of algebraic operations and quadratures of some functions.

There are two main techniques in the process of solving the system:

\begin{itemize}
\item{}  Determination  of constants of motion:
Constants of motion provide us
 foliations such that $\Gamma$ is tangent to the leaves of the foliation,
 and reducing in this way the problem to a  family of  lower dimensional problems,  one
 on each  leaf.
 \item{}  Search for symmetries of the vector field:
 The knowledge of  infinitesimal one-parameter groups of symmetries of the vector field (i.e.
  of the system of differential equations), suggests us to use  {adapted local coordinates},  the system  decoupling then
  into  lower dimensional subsystems.
\end{itemize}

   More specifically,  the knowledge of $r$  functionally independent  (i.e. such that $dF_1\wedge\cdots\wedge dF_r\ne 0$)
 constants of motion,
$F_1,\ldots,F_r$,
allows us to
reduce the problem to that of a  family of vector fields
 $\widetilde \Gamma_c$  defined in the  $N-r$ dimensional submanifolds  $M_c$ given by  the
level sets of the vector function of rank $r$,
  $(F_1,\ldots,F_r):M\to \mathbb{R}^r$.
   Of course the  best situation is when $r=N-1$: the leaves
are one-dimensional, giving us the  solutions to the problem, up to a reparametrisation.

There is another way of reducing the problem. Given an  infinitesimal  symmetry (i.e. a
vector field  $X$ such
that  $[X,\Gamma]=0$), then, according to the Straightening out Theorem \cite{AbrahamMar,AbrahamMarRat, Crampinbook},  in a neighbourhood of a point where $X$ is different from zero we can  choose adapted coordinates,
$(y^1,\ldots,y^N)$,
for which  $X$ is written as
$$X=\pd{}{y^N}\ .$$

Then, the symmetry condition    $[X,\Gamma]=0$ implies that $\Gamma$ has the form
$$
\Gamma=\bar f^1(y^1,\ldots,y^{N-1})\,\pd{}{y^1}+\ldots +\bar f^{N-1}(y^1,\ldots,y^{N-1})\,\pd{}{y^{N-1}}+
\bar f^N(y^1,\ldots,y^{N-1})\pd{}{y^N}\ ,
$$
 and its  integral curves are obtained by solving the system of differential equations
$$\left\{\begin{array}{ccl}
{\displaystyle \frac{dy^i}{dt}}&=&\bar f^i(y^1,\ldots,y^{N-1})\ ,\qquad
 i=1,\ldots ,N-1\cr
{\displaystyle \frac{dy^N}{dt}}&=& \bar f^N(y^1,\ldots,y^{N-1}).
\end{array}
\right.
$$
We have  reduced the problem to a  {subsystem involving only the first
 $N-1$ equations}, and once this has been solved,  the last equation is used to obtain  the function
 $y^N(t)$  by means of one more quadrature.

 Note  that the new coordinates, $y^1,\ldots,y^{N-1}$, are such that $Xy^1=\cdots=Xy^{N-1}=0$, i.e. they  are constants of the motion for $X$
 and therefore we cannot easily  find such coordinates in a general case.

Moreover, the information provided by
  two different symmetry vector fields cannot be used simultaneously in the general case,
because it is not possible to find local coordinates
 $(y^1,\ldots,y^N)$  such that
 $$X_1=\pd{}{y^{N-1}}\ ,\qquad X_2=\pd{}{y^N}\ ,
$$
 unless that  $[X_1,X_2]=0$.

 In terms of adapted coordinates for  the dynamical vector field $\Gamma$, i.e. $\Gamma=\partial/\partial y^N$,   the integration is immediate, the solution curves being given by
 $$y^k(t)=y^k_0, \quad k=1,\ldots ,N-1,\qquad y^N(t)=y^N(0)+t.
 $$
 This proves that  the concept of integrability by quadratures depends on the choice of initial coordinates, because in these adapted coordinates the system is easily solved.

 However, it will be proved that when $\Gamma$ is part of a family of vector fields satisfying appropriate conditions, then
 it is  integrable  by quadratures for  any choice of initial coordinates.

 Both,  constants of motion and infinitesimal symmetries, can be used simultaneously if some compatibility conditions are satisfied.
We can say that a system admitting $r<N-1$ functionally independent constants of motion, $F_1,\ldots,F_r$,  is integrable when we know
 furthermore $s$ commuting   infinitesimal symmetries $X_1,\ldots,X_s$, with $r+s=N$  such that
$$[X_a,X_b]=0, \ a,b=1,\ldots, s,\qquad
\textrm{and}\qquad X_aF_\alpha=0, \quad \forall a=1,\ldots,s, \alpha=1,\ldots r.
$$

The  constants of motion determine a $s$-dimensional  foliation  (with $s=N-r$) and the former condition
means that  the restriction of the $s$ vector fields $X_a$ to the leaves are tangent to such leaves.

Sometimes we have additional geometric structures that are  compatible with the dynamics.
For instance, a  symplectic structure  $\omega$ on a $2n$-dimensional manifold $M$.
Such a 2-form  relates, by contraction,   in a one-to-one way, vector fields and 1-forms.
Vector fields $X_F$ associated with exact 1-forms  $dF$ are said to be  Hamiltonian vector fields. 
We say that $\omega$ is compatible means that the dynamical vector field itself is a Hamiltonian vector field  $X_H$.

Particularly interesting is the Arnold--Liouville  definition of (Abelian)  complete integrability ($r=s=n$, with $N=2n$) \cite{arnold,arnoldkozlov,K83,JLiou53}.
The vector fields are $X_a=X_{F_a}$ and, for instance, $F_1=H$.

The regular Poisson bracket defined by $\omega$ (i.e. $\{F_1,F_2\}=X_{F_2}F_1$), allows us to express the above  tangency conditions as
$X_{F_b}F_a=\{F_a,F_b\}=0$
-- i.e. the $n$ functions are constants of motion in involution and their corresponding Hamiltonian vector fields commute.

Our aim  is to study  integrability in absence of additional compatible structures, the main tool being properties of Lie algebras of vector fields containing  the given vector field,
 very much in the approach started by Lie.

 The problem of integrability by quadratures depends on the determination by quadratures of the necessary first-integrals and
  on finding adapted coordinates, or,  in other words, in finding a sufficient number of invariant tensors.

The set  $ \mathfrak{X}_\Gamma(M)$ of  strict infinitesimal symmetries of
 $\Gamma\in \mathfrak{X}(M)$  is a linear space:
 $$ \mathfrak{X}_\Gamma(M)=\{X\in \mathfrak{X}(M)\mid [X,\Gamma]=0\}\ .
$$
The  flow of vector fields
 $X\in \mathfrak{X}_\Gamma(M)$  preserve the
set of integral  curves of   $\Gamma$.

The set of vector fields generating flows  preserving the set of integral
curves of $\Gamma$ up to a reparametrisation is
a real
linear space containing $\mathfrak{X}_\Gamma(M)$ and will be denoted
$$ \mathfrak{X}^\Gamma(M)=\{X\in \mathfrak{X}(M)\mid [X,\Gamma]=f_X\, \Gamma\}\ ,\quad
f_X\in C^\infty (M). $$
The flows of vector fields in $\mathfrak{X}^\Gamma(M)$
 preserve the one-dimensional distribution generated by $\Gamma$. Moreover, for any function $f\in C^\infty (M)$,
 $\mathfrak{X}^\Gamma(M)\subset \mathfrak{X}^{f\Gamma}(M)$, i.e. $\mathfrak{X}^\Gamma(M)$ only depends
 of the distribution generated by $\Gamma$ and not on $\Gamma$ itself.

One can check that $\mathfrak{X}^\Gamma(M)$  is a real Lie algebra and
$\mathfrak{X}_\Gamma(M)$  is a Lie subalgebra of $\mathfrak{X}^\Gamma(M)$.
However  $\mathfrak{X}_\Gamma(M)$ is not an ideal in $\mathfrak{X}^\Gamma(M)$.

As indicated above, finding constants of motion for $\Gamma$ is not an easy task, at least in absence of a compatible symplectic structure.
However,  the explicit knowledge of first integrals of a given dynamical system has
proved to be of great importance in the study of the qualitative properties of
the system.   The important point is that an appropriate  set of infinitesimal symmetries of $\Gamma$ can also provide constants of motion. More specifically,
let $\{X_1,\ldots,X_d\} $ be a set of $d$ vector fields taking  linearly independent values in every point and
 which are infinitesimal symmetries of $\Gamma$. If they generate   an involutive distribution, i.e. there exist functions $f_{ij}\,^k$ such that $[X_i,X_j]=f_{ij}\,^kX_k$,
then, for each triple of numbers $i,j,k$ the functions
$f_{ij}\,^k$  are constants of the motion, i.e. $\Gamma (f_{ij}\,^k)=0$.
In fact, Jacobi identity for the vector fields $\Gamma,X_i,X_j$, i.e.
$$[[\Gamma,X_i],X_j]+[[X_i,X_j],\Gamma]+ [[X_j,\Gamma],X_i]=0,$$ leads to
$$[[X_i,X_j],\Gamma]=0\Longrightarrow [f_{ij}\,^kX_k,\Gamma]=-\Gamma(f_{ij}\,^k)\, X_k=0.
$$
Moreover, for any other index $l$, $X_l(f_{ij}\,^k)$ is also a constant of motion, because as $X_l$ is a symmetry of $\Gamma$,
then $\mathcal{L}_{\Gamma}\left(\mathcal{L}_{X_l}(f_{ij}\,^k)\right)=
\mathcal{L}_{X_l}\left(\mathcal{L}_{\Gamma}(f_{ij}\,^k)\right)=0$.

The constants of motion so obtained are not functionally independent but at least this proves the usefulness of finding these families of vector fields when looking for constants of motion. This points out the convenience of extending the theory from Lie algebras of symmetries to
involutive distributions, as we will do in the final part of the paper.

\section{Lie theorem of integrability by quadratures}

The first important result is due to Lie who established the following theorem:

\begin{theorem}
If $n$ vector fields, $X_1$,\ldots,$X_n$, which are linearly independent in each point of  an open set  $U\subset\mathbb{R}^n$, generate a  {solvable Lie algebra} and  are   such that $[X_1,X_i]=\lambda_i\, X_1$ with $\lambda_i\in \mathbb{R} $,
then  the differential equation  $\dot x=X_1(x)$ is solvable by quadratures in $U$.
\end{theorem}

We only prove the  simplest case $n=2$.  The differential equation can be integrated if we are able to find a first integral $F$ (i.e. $X_1F=0$),   such that $dF\ne 0$ in $U$.  The straightening out theorem \cite{AbrahamMar,AbrahamMarRat, Crampinbook}, says that such
a function $F$  locally exists.
$F$ implicitly defines one variable, for instance $x_2$,
  in  terms of the other one by $F(x_1,\phi(x_1))=k$.

   If  $X_1$ and $X_2$  are  such that $[X_1,X_2]=\lambda_2\, X_1$, and  $\alpha_0$ is  a 1-form,
   defined up to multiplication by a function,  such that $i(X_1)\alpha_0=0$, as $X_2$ is linearly independent of $X_1$ at each point, $i(X_2)\alpha_0\ne 0$, and we can
see that the
1-form $\alpha=(i(X_2)\alpha_0)^{-1}\alpha_0$ is such that $i(X_1)\alpha=0$ and satisfies, by construction,  the condition $i(X_2)\alpha=1 $. Such 1-form $\alpha$  is  closed, because
 $X_1$ and $X_2$ generate $\mathfrak{X}(\mathbb{R}^2)$ and
$$ d\alpha(X_1,X_2)=X_1\alpha(X_2) -X_2\alpha(X_1)+\alpha([X_1,X_2])= \alpha([X_1,X_2])=\lambda_2\, \alpha(X_1)=0.
$$
Therefore, there exists, at least locally, a function $F$ such that $\alpha=dF$, and it is
given by
$$F(x_1,x_2) =\int_\gamma\alpha,
$$
where $\gamma$ is any curve with end in the point $(x_1,x_2) $. This  is the function we were looking for, because $dF=\alpha$ and then
$$i(X_1)\alpha=0\Longleftrightarrow X_1F=0,\qquad i(X_2)\alpha=1\Longleftrightarrow X_2F=1.
$$
We do not present here the proof for general $n$ because it appears as a particular case of the more general situation we consider later on.
The result of this theorem has been slightly generalized for adjoint-split solvable Lie algebras in \cite{K05}.

\section{Recalling some basic concepts of cohomology}

Let  be $\mathfrak{g}$   a Lie algebra and  $\mathfrak{a}$ a $\mathfrak{g}$-module, or in other words,
 $\mathfrak{a}$ is a linear space  that is carrier space for a linear
representation $\Psi$ of  $\mathfrak{g}$, i.e.   $\Psi \colon \mathfrak{g} \to \textrm{End\,} \mathfrak{a}$ satisfies
$$\Psi (a) \Psi (b)-\Psi (b) \Psi(a)=\Psi ([a,b]),\quad \forall a,b\in \mathfrak{g}.$$

By a   $k$-cochain  we mean a $k$-linear alternating
map $\alpha:\mathfrak{g}\times\cdots\times\mathfrak{g}\to \mathfrak{a}$.
If $C^k(\mathfrak{g},\mathfrak{a})$ denotes
the linear space of $k$-cochains,
for each $k\in\mathbb{N}$ we  define
$\delta_k:C^k(\mathfrak{g},\mathfrak{a})\to
C^{k+1}(\mathfrak{g},\mathfrak{a})$ by (see e.g. \cite{CE48} and \cite{CI88} and references therein)
$$\begin{array}{rcl}
(\delta_k\alpha)(a_1,\dots,a_{k+1})
 &= &{\displaystyle\sum_{i=1}^{k+1} (-1)^{i+1} \Psi(a_i)
        \alpha(a_1,\dots,\widehat a_i,\dots,a_{k+1})+ } \\
&+& {\displaystyle\sum_{i<j} (-1)^{i+j}
\alpha([a_i,a_j],a_1,\dots,\widehat
                a_i,\dots,\widehat a_j,\dots,a_{k+1})},
\end{array}
$$
where $\widehat a_i$ denotes, as usual, that the element $a_i\in \mathfrak{g}$ is omitted.

  The  linear maps $\delta _k$ can be shown to satisfy  {$\delta _{k+1}\circ
\delta _k=0$}, and consequently
the linear operator $\delta$ on
$C(\mathfrak{g},\mathfrak{a}) = \bigoplus_{k =0}^\infty                                                                                                                                                                                                                                                                         
C^k(\mathfrak{g},\mathfrak{a})$ whose restriction to
each $C^k(\mathfrak{g},\mathfrak{a})$ is $\delta_k$, satisfies $\delta^2 = 0$.  We will then
denote
$$\begin{array}{rcl}
B^k(\mathfrak{g},\mathfrak{a}) &=& \{\alpha \in C^k(\mathfrak{g},\mathfrak{a}) \mid \exists\beta\in
C^{k-1}(\mathfrak{g},\mathfrak{a}) \text{ such that }\alpha = \delta \beta \}
  = \textrm{Image\,} \delta_{k-1},  \\&&\\
Z^k(\mathfrak{g},\mathfrak{a}) &= &\{\alpha\in C^k(\mathfrak{g},\mathfrak{a}) \mid \delta\alpha = 0\} = \ker
\delta_k.
\end{array}
$$
The elements of $Z^k(\mathfrak{g},\mathfrak{a})$ are called  $k$-cocycles, and those of
$B^k(\mathfrak{g},\mathfrak{a})$ are called  $k$-cobound\-aries.
As  $\delta$ is such that $\delta^2 = 0$, we see that
$B^k (\mathfrak{g},\mathfrak{a})\subset Z^k(\mathfrak{g},\mathfrak{a})$. The  {$k$-th cohomology group} $H^k(\mathfrak{g},\mathfrak{a})$ is
$$
H^k(\mathfrak{g},\mathfrak{a}) := \frac{Z^k(\mathfrak{g},\mathfrak{a})}{B^k(\mathfrak{g},\mathfrak{a})} \,,
$$
and we will define $B^0(\mathfrak{g},\mathfrak{a})=0$, by convention.

We are interested in the case  where  $\mathfrak{g}$ is  a finite-dimensional Lie subalgebra of $\mathfrak{X}(M)$, $\mathfrak{a}=\bigwedge^p(M)$,
 and consider the action of $\mathfrak{g}$  on $\mathfrak{a}$  given by $\Psi(X)\zeta=\mathcal{L}_{X}\zeta$.
The case $p=0$,  has been used, for instance,
 in the study of  weakly invariant differential equations as shown in \cite{COW93}.
 The cases   $p=1,2,$ are also interesting  in mechanics \cite{CI88}.

  Coming back to the particular case $p=0$,  $\mathfrak{a}=\bigwedge^0(M)=C^\infty(M)$, $\mathfrak{g}= \mathfrak{X}(M)$, the elements of $Z^1(\mathfrak{g},\bigwedge^0(M))$ are linear maps
 $h:\mathfrak{g}\to C^\infty(M)$ satisfying
 $$
 (\delta_1 h)(X,Y)=\mathcal{L}_{X} h(Y) - \mathcal{L}_{Y} h(X) - h([X,Y])=0\ ,\qquad X,Y\in \mathfrak{X}(M),
 $$
and those of $B^1(\mathfrak{g},C^\infty(M))$ are linear maps  $h:\mathfrak{g}\to C^\infty(M)$  for which
 $\exists g\in C^\infty(M)$ with
 $$h(X) = \mathcal{L}_{X}g\ .$$
{\bf Lemma} {\it
Let  $\{X_1,\ldots,X_n\}$ be a set of $n$ vector fields whose values are linearly independent at each point of an  $n$-dimensional manifold $M$. Then:

1)  The necessary and sufficient condition for the system of equations for $f\in C^\infty(M)$
$$
  X_i f = h_i, \qquad h_i\in C^\infty(M) ,\quad i=1,\dots,n,
$$
to have a solution is that the 1-form $\alpha\in \bigwedge ^1(M)$ such that
$\alpha(X_i)=h_i$ be an exact 1-form.

2)   If the previous $n$ vector fields generate a $n$-dimensional real  Lie algebra $\mathfrak{g}$ (i.e. there exist real numbers $c_{ij}\,^k$ such that $[X_i,X_j]=c_{ij}\,^k\, X_k$), then
 the necessary condition for the system of equations to have a solution is that the
 $\mathbb{R}$-linear function
 $h:\mathfrak{g}\to C^\infty(M)$  defined by $h(X_i)=h_i$ is a 1-cochain  that is a 1-cocycle.}

{\sl Proof}.-  1)  For any pair of indices $i,j$, if $X_i f = h_i$ and $X_j f = h_j$,
then, as  $\exists f_{ij}\,^k\in C^\infty(M)$ such that $[X_i,X_j]=f_{ij}\,^k\, X_k$,
$$X_i(X_j f)-X_j (X_i f)=[X_i,X_j]f=f_{ij}\,^k\, X_kf\,\Longrightarrow X_i(h_j)-X_j(h_i)
-f_{ij}\,^k\, h_k=0,
$$
and as  $\alpha(X_i)=h_i$, we obtain that as
$$d\alpha(X_i,X_j)=X_i\alpha(X_j)-X_j\alpha(X_i)-\alpha([X_i,X_j])=X_i(h_j)-X_j(h_i)
-f_{ij}\,^k\, h_k,
$$
the 1-form $\alpha$ is closed. Consequently,
 a necessary condition for the existence of the solution of the system  is that $\alpha$ be closed.

2) Consider $\mathfrak{a}=C^\infty(M)$,  $\mathfrak{g}$ the $n$-dimensional real  Lie algebra generated by the vector fields $X_i$,  and  the cochain determined by the linear map $h:\mathfrak{g}\to C^\infty(M)$. Now the necessary condition for the existence of the solution is written as:
$$
X_i(h_j)-X_j(h_i)-c_{ij}\,^k\, h_k=(\delta_1 h)(X_i,X_j)=0.
$$
This is just the 1-cocycle condition.

Most properties of differential equations are of a  {local character}: closed forms are
locally exact and   we can restrict ourselves to appropriate open subsets $U$ of $M$, i.e. open submanifolds, where the closed 1-form is exact, .
Then if $\alpha$ is closed,  it is locally exact, $\alpha=df$ in a certain open $U$, $f\in C^\infty(U)$,  and the solution of the system can be found  by one quadrature:
the solution  function $f$ is  given by   the quadrature
$$
  f(x)=\int_{\gamma_x}\alpha,
$$
where $\gamma_x$ is any path joining some  reference point $x_0\in U$ with $x\in U$.

We also remark that
 $\alpha$ is exact, $\alpha=df$, if
and only if  $\alpha(X_i)=df(X_i)=X_if=h_i$, i.e. $h$ is a coboundary,  $h=\delta f$.

 In the particular case of the appearing functions $h_i$   being constant
the condition for the existence of local solution reduces to $\alpha([X,Y])=0$, for
each pair of elements, $X$ and $Y$ in  $\mathfrak{g}$, i.e. $\alpha$ vanishes on the derived Lie algebra $\mathfrak{g}'=[\mathfrak{g},\mathfrak{g}]$.
In particular when  $\mathfrak{g}$ is Abelian there is not any condition.

\section{A generalisation of Lie theory of integration}

Consider a family of  $N$ vector fields,  $X_1,\dots,X_N$, defined on
a $N$-dimensional manifold $M$ and  assume that they  close a Lie algebra $L$ over the real
numbers
$$
 [X_i,X_j] = c_{ij}\,^k \,X_k \,, {\quad} i,j,k = 1,\dots,N,
$$
and that, in addition,  they  {span a basis of $T_xM$ at  every point $x\in M$}.
We pick up an element in the family,   $X_1$, the dynamical vector field.
To emphasize its
special r\^ole we will often denote it by  {$\Gamma\equiv X_1$}.

Our goal, is to  obtain the integral curves $\Phi_t:M\rightarrow M$
of $\Gamma$
$$
 (\Gamma f) (\Phi_t(x)) = \frac{d}{dt} f(\Phi_t(x)),\quad \forall f\in C^\infty (x),\ x\in M,
$$
 by using quadratures
(operations of integration, elimination and partial differentiation).
The number of quadratures is given by the number of
integrals of known functions depending on a finite number of parameters,
that are performed.
$\Gamma$ plays a distinguished r\^ole since it represents the dynamics to be integrated.

Our  approach is
concerned with the construction of a sequence of nested Lie subalgebras $L_{\Gamma,k}$ of the Lie algebra $L$,  and it will be essential that  $\Gamma$
belongs to all these subalgebras.  This construction, for which more details can be found in \cite{CFGR15},  will be carried out  in several steps.

 The first one will be to  {reduce, by one quadrature}, the original problem to a similar one but  {with a Lie subalgebra $L_{\Gamma,1}$} of the Lie algebra $L$
(with $\Gamma\in L_{\Gamma,1}$)
whose elements span at every point the tangent space of the
leaves of a certain foliation.

 If iterating the procedure we end up with an Abelian Lie algebra
we can, with one more quadrature, obtain the flow of the dynamical
vector field.

We determine the foliation through a family of functions
that are constant on the leaves. We first  {consider the ideal} in $L$
$$
 L_{\Gamma,1} = \langle \Gamma\rangle + [L,L] \,,{\quad} \dim L_{\Gamma,1} = n_1,
$$
that, in order to make the notation simpler, we will assume to be generated
by the first $n_1$ vector fields of the family (i.e. $L_{\Gamma,1}=\langle \Gamma,X_2,\dots,
X_{n_1}\rangle$). This can always be achieved by choosing appropriately the
basis of $L$.

Now  take $\zeta_1$ in the annihilator\ of $L_{\Gamma,1}$,  i.e. $\zeta_1$ is in the set $L_{\Gamma,1}^0$ made up by the
elements of  $L^*$ killing\ vectors of	 $L_{\Gamma,1}$,
and  {define the 1-form $\alpha_{\zeta_1}$  on  $M$ by} its action on
the vector fields in $L$:
$$\alpha_{\zeta_1}(X)=\zeta_1(X),\quad\mathrm{for}\ X\in L.$$
As $\alpha_{\zeta_1}(X)$ is a constant function on $M$, for any vector
field in $L$, we have
$$d\alpha_{\zeta_1}(X,Y)=\alpha_{\zeta_1}([X,Y])=\zeta_1([X,Y])=0,\quad\mathrm{for}\  X,Y\in L,\zeta_1\in L_{\Gamma,1}^0.$$
Therefore  {the 1-form $\alpha_{\zeta_1}$ is closed} and by application of the result of the lemma
 {the system of partial differential equations}
$$
 X_i Q_{\zeta_1} =\alpha_{\zeta_1}(X_i),\quad i=1,\dots,n, \quad Q_{\zeta_1}\in C^\infty(M),
$$
 has a unique (up to the addition of a constant) local solution
which can be obtained by one quadrature. Moreover, if
we fix the same
reference point $x_0$ for any $\zeta_1$,
 {$\alpha_{\zeta_1}$ depends linearly on $\zeta_1$} and, if $\gamma_x$ is independent of $\zeta_1$, we have that the correspondence
$$L_{\Gamma,1}^0\ni\zeta_1\mapsto Q_{\zeta_1}\in C^\infty(M)$$
defines an injective linear map.

The system expresses that  {the vector fields
in $L_{\Gamma,1}$ (including $\Gamma$) are tangent to
$$
  N_1^{[Y_1]}=\{x\mid Q_{\zeta_1}(x)=\zeta_1(Y_1),\,\zeta_1\in L_{\Gamma,1}^0\}\subset M
$$
for any $[Y_1]\in L/L_{\Gamma,1}$}. Locally, for an open
neigbourhood $U$, the $N_1^{[Y_1]}$'s define a smooth foliation of
$n_1$-dimensional leaves.

Now, we  repeat the previous procedure by taking $L_{\Gamma,1}$ as the  Lie algebra and
any leaf $N_1^{[Y_1]}$ as the manifold.
 The new subalgebra $L_{\Gamma,2}\subset L_{\Gamma,1}$ is defined by
$$
 L_{\Gamma,2} = \langle \Gamma\rangle + [L_{\Gamma,1},L_{\Gamma,1}] \,,{\quad} \dim L_{\Gamma,2} = n_2\,,
$$
and taking $\zeta_2\in L_{\Gamma,2}^0\subset L_{\Gamma,1}^*$ (the annihilator of $L_{\Gamma,2}$),
 we arrive at a new system of partial differential equations
$$
 X_i Q_{\zeta_2}^{[Y_1]} =\zeta_2(X_i),\quad i=1,\dots,n_1, \quad
 Q_{\zeta_2}^{[Y_1]}\in C^\infty(N_1^{[Y_1]}) \,,
$$
that can be solved with one quadrature and  such  $Q_{\zeta_2}^{[Y_1]}$
depends linearly on $\zeta_2$.

It will be useful to extend
$Q_{\zeta_2}^{[Y_1]}$ to $U$. We
first introduce the map
$$U\ni x\mapsto [Y_1^{^x}]\in L_{\Gamma,0}/L_{\Gamma,1}$$
where $x$ and $[Y_1^{^x}]$ are  related by
the equation $Q_{\zeta_1}(x)=\zeta_1(Y_1^{^x})$, that correctly
determines the map.
Now, we define $Q_{\zeta_2}\in C^\infty(U)$ by
$Q_{\zeta_2}(x)= Q_{\zeta_2}^{[Y_1^{^x}]}(x)$.
Note that by construction $x\in N_1^{[Y^{^x}_1]}$
and, therefore the definition makes sense.
The resulting function
$Q_{\zeta_2}(x)$
is smooth provided the reference point of the lemma changes
smoothly from leave to leave.

 The construction is then iterated by defining
$$N_2^{[Y_1][Y_2]}=\{x\mid Q_{\zeta_1}(x)=\zeta_1(Y_1), \quad Q_{\zeta_2}(x)=\zeta_2(Y_2),\
{\rm with}\
\zeta_1\in L_{\Gamma,1}^0, \zeta_2\in L_{\Gamma,2}^0\}\subset M,$$
for $[Y_1]\in L_{\Gamma,0}/L_{\Gamma,1}$ and $[Y_2]\in L_{\Gamma,1}/L_{\Gamma,2}$.
Note that $L_{\Gamma,2}$ generates at every point the tangent space
of $N_2^{[Y_1][Y_2]}$, therefore we can proceed as before.

 The algorithm ends if after some steps, say $k$, the Lie algebra
$L_{\Gamma,k}=\langle X_1,\dots,X_{n_k}\rangle$,
whose vector fields are tangent to the $n_k$-dimensional
leaf $N_k^{[Y_1],\dots,[Y_k]}$, is Abelian.
In this moment the system of equations
$$
  X_i Q_{\zeta_k}^{[Y_1],\dots,[Y_k]}=\zeta_k(X_i),\quad i=1,\dots,n_{k-1},\quad
Q_{\zeta_k}^{[Y_1],\dots,[Y_k]}\in C^\infty(N_k^{[Y_1],\dots,[Y_k]}),
$$
can be solved locally by one more quadrature for any  $\zeta_k\in L_{\Gamma,k}^*$.

Remark that,  as the Lie algebra $L_{\Gamma,k}$ is Abelian, the integrability condition is always satisfied
and we can take $\zeta_k$ in the whole of $L_{\Gamma,k}^*$ instead of $L_{\Gamma,k}^0$.
Then, as before, we extend
the solutions to $U$ and call them $Q_{\zeta_k}$.

With all these ingredients  {we can find the flow of $\Gamma$ by performing
only algebraic operations}. In fact, consider the formal direct sum
$$\Xi=L_{\Gamma,1}^0\oplus L_{\Gamma,2}^0\oplus\cdots\oplus L_{\Gamma,{k}}^0\oplus L_{\Gamma,k}^*,$$
that, as one can check, has dimension $n$.

The linear
maps $L_{\Gamma,i}^0\ni\zeta_i\mapsto Q_{\zeta_i}\in C^\infty(U)$
can be extended to $\Xi$ so that to any
$\xi\in\Xi$ we assign a
$Q_\xi\in C^\infty(U)$.
Now  {consider a basis
$$\{\xi_1,\dots,\xi_n\}\subset\Xi.$$}
 {The associated functions $Q_{\xi_j},j=1,\dots,n$ are independent and satisfy
$$
 \Gamma Q_{\xi_j}(x)= \xi_j(\Gamma) \,,{\quad}  j=1,2,\dots,n,
$$}
where it should be noticed that as $\Gamma\in L_{\Gamma,l}$ for any
$l=0,\dots,k$, the right hand side is well defined, and
 we see from here  that
in the coordinates given by the $Q_{\xi_j}$'s
the vector field $\Gamma$ has constant components and, then,
it is trivially integrated  {
 $$Q_{\xi_j}(\Phi_t(x))=Q_{\xi_j}(x)+ \xi_j(\Gamma) t.$$}
Now, with algebraic operations, one can derive the flow $\Phi_t(x)$.
Altogether we have performed $k+1$ quadratures.

\section{Algebraic properties}
The previous procedure   works if it reaches an end point (i.e.
if there is a smallest non negative integer $k>0$ such that
$$
 L_{\Gamma,{k}}=\langle \Gamma\rangle+[L_{\Gamma,{k-1}},L_{\Gamma,{k-1}}]\,,
$$
is an Abelian algebra).  In that case we say that $(M,L,\Gamma)$ is
Lie integrable of order $k+1$.

The content of the previous section can, thus, be  summarized in the following

\begin{proposition}  If $(M,L,\Gamma)$ is Lie integrable of order $r$,
then  the integral curves of $\Gamma$ can be obtained by $r$ quadratures.
\end{proposition}

We will discuss below some necessary and sufficient conditions for
the Lie integrability.

\begin{proposition}
  If $(M,L,\Gamma)$ is Lie integrable then $L$ is solvable.
  \end{proposition}

{\sl Proof}.-  Let $L_{(i)}$ be the elements of the derived series,
$L_{(i+1)}=[L_{(i)},L_{(i)}]$, $L_{(0)}=L$, (note that $L_{(i)}=L_{0,i}$).
Then,
$$
  L_{(i)}\subset L_{\Gamma,i},
$$
and if the system is Lie integrable  (i.e.  $L_{\Gamma,k}$ is Abelian
for some $k$), then we have $L_{(k+1)}=0$ and, therefore, $L$ is solvable.

\begin{proposition}
  If $L$ is solvable and $A$ is an Abelian ideal of $L$,
then $(M,L,\Gamma)$ is Lie integrable for any $\Gamma\in A$.
\end{proposition}

{\sl Proof}.-  Using that
$A$ is an ideal containing $\Gamma$, we can show that
$$A+L_{\Gamma,i}=A+L_{(i)}.$$
We proceed again by induction: if the previous holds, then
$$
\begin{array}{rcl}
A+L_{\Gamma,i+1}&=&A+[L_{\Gamma,i},L_{\Gamma,i}]=A+[A+L_{\Gamma,i},A+L_{\Gamma,i}]=\cr
&=&A+[A+L_{(i)},A+L_{(i)}]=A+L_{(i+1)}.
\end{array}
$$
Now $L$ is solvable if some $L_{(k)}=0$ and therefore $L_{\Gamma,k}\subset A$,
i.e. it is Abelian and henceforth the system is Lie integrable.
Note that the particular case
$A=\langle \Gamma\rangle$
corresponds to the standard Lie theorem.

Nilpotent algebras of vector fields  also play an interesting role in the
integrability of vector fields.

\begin{proposition}
  {If $L$ is nilpotent,  $(M,L,\Gamma)$ is Lie integrable for any $\Gamma\in L$}.
  \end{proposition}

{\sl Proof}.- Let us consider the central series
$L^{(i+1)}=[L,L^{(i)}]$ with $L^{(0)}=L$. Now,
$L$ nilpotent means that there is a $k$ such that $L^{(k)}=0$.
 It is easy to see, by induction, that
$L_{\Gamma,i}\subset \langle \Gamma\rangle +L^{(i)}$
and therefore $L_{\Gamma,k}= \langle \Gamma\rangle$
 is Abelian and the system is Lie integrable.

From the previous propositions, we can derive the following
\begin{corollary}\label{cor1}
Let $(M,L,\Gamma)$ be Lie integrable of order $r$. Then:

\textrm{(a)} If $r_s$ is
the minimum positive integer such that $L_{(r_s)}=0$, then
$r\geq r_s.$

\textrm{(b)} If $L$ is nilpotent $r_n$ is the smallest natural number
such that $L^{(r_n)}=0$,
$r\leq r_n.$
\end{corollary}

\section{An interesting example}

We now analyse the particular case of a recently studied superintegrable system \cite{CCR13},  where we dealt with an example of a potential that is not separable
 but is superintegrable with high order first integrals \cite{PW11}, by studying limits of some  potentials   related to Holt potential \cite{H82}.
 Even if the system is Hamiltonian, that is, the dynamical vector field $\Gamma=X_H$ is obtained from a Hamiltonian function $H$ by making use of a symplectic structure $\omega_0$ defined in a cotangent bundle $T^*Q$
 we deliberately forget this fact and analyse the situation
by simply considering this system just as a dynamical system (without mentioning the existence of a symplectic structure) and focusing our attention on the Lie algebra structure of the symmetries.

Suppose that the dynamics is given by the  vector field $\Gamma=X_1$ defined in $M=\mathbb{R}^2\times\mathbb{R}^2$ with coordinates $(x,y,p_x,p_y)$ given by
$$
 \Gamma=X_1= p_x\pd{}{x} + p_y\pd{}y  - \frac{k_2}{y^{ 2/3}}\pd{}{p_x}
  + \frac{2}{3} \frac{k_2\,x+k_3}{y^{ 5/3}} \pd{}{p_y} \,,
$$
where $k_2$ and $k_3$ are arbitrary constants.

Consider in this case
 the following three vector fields:
$$
\begin{array}{rcl}
X_2&=& {\displaystyle
\left(6\, p_x^2+3\, p_y^2+k_2\frac {6x}{y^{ 2/3}} + k_3 \frac {6}{y^{ 2/3}}\right)\pd{}{x}+(6\, p_xp_y+9\, k_2y^{ 1/3})\pd{}y}  \\
&-&\displaystyle{k_2\frac {6}{y^{ 2/3}}\, p_x
\pd{}{p_x}+\left(4k_2\frac {x}{y^{ 5/3}}-3\frac {1}{y^{ 2/3}}\,p_y\right)\pd{}{p_y}}   \,,
\end{array}
$$
$$
\begin{array}{rcl}
X_3&=&{\displaystyle\left(4\, p_x^3+4\, p_xp_y^2+\frac{8(k_2x+k_3)}{y^{ 2/3}}p_x+12k_2\, y^{ 1/3}\,p_y\right)
\pd{}{x}}\\
&+&\left(4p_x^2\, p_y+12k_2\, y^{ 1/3}\,p_x\right)\displaystyle{\pd{}{y}}
- {4k_2\frac {1}{y^{ 2/3}}p_x^2\,\pd{}{p_x}}  \\
&+&\left(\displaystyle{ \frac{8}{3} \frac{k_2x+k_3}{y^{5/3}} p_x^2}
- 4k_2 \frac{1}{y^{ 2/3}} p_xp_y - 12\,k_2^2 \frac{1}{y^{1/3}}\right)
\displaystyle{\pd{}{p_y} }  \,,
\end{array}
$$
and
$$\begin{array}{rcl}
X_4&=&{\displaystyle\left(6p_x^5+12\, p_x^3p_y^2+24 \frac{k_3 + k_2 x}{y^{2/3}}p_x^3+108\,k_2 y^{1/3} p_x^2 p_y  +324\, k_2^2   y^{2/3} p_x\right)\pd{}{x}}\\&+&
{\displaystyle \left(6\,  p_x^4 p_y+36\,k_2 y^{1/3} p_x^3   \right)\pd{}y-6\, \left(\frac{ k_2 }{y^{2/3}} p_x^4-972 k_2^3 \right)\pd{}{p_x}}
\cr&+&{\displaystyle \left(4\,\frac{k_3 + k_2 x}{y^{5/3}}p_x^4-12\frac{ k_2}{ y^{2/3}}-108\, k_2^2\frac 1{y^{1/3}}p_x^2 \right)\pd{}{p_y}} \,.
\end{array}
$$

In order to apply the theory developed above, it suffices to compute the commutation relations among the fields:
\begin{equation}
 [X_2, X_3]=0 \,,{\qquad}
 [X_2, X_4] = 1944\, k_2^3\,\Gamma\,,{\qquad}
 [X_3, X_4] = 432\, k_2^3 \,X_2
\end{equation}
together with:
\begin{equation}
[X_1 , X_i]=0 \,,{\quad} i=2,3,4.
\end{equation}

Therefore, $\Gamma$ and the three
vector fields $X_2,X_3,X_4$ generate a four-dimensional real Lie algebra ${L}$, whose center is  generated by $\Gamma=X_1$.
The derived algebra ${L}_{(1)}\subset {L}$  is two-dimensional and
it is generated by $X_1$ and $X_2$, i.e. ${L}_{(1)}$ is Abelian.
Finally,   the second derived
algebra   ${L}_{(2)}$ reduces to the trivial algebra, because  ${L}_{(1)} $ is Abelian. That is,
${L}_{(2)}=[{L}_{(1)},{L}_{(1)} ]=\{0\}$

In summary,   the Lie algebra ${L}$ is
solvable with solvability index $r_s=2$. However,
${L}^{(2)}=[{L},{L}_{(1)} ]$ is not trivial but ${L}_{(1)} $ is  the one-dimensional ideal in ${L}$
 generated by $X_1$, and this implies that the Lie algebra is nilpotent with $r_n=3$.

 According to the previous results, we can conclude that
$(M,L,\Gamma)$   is Lie integrable for any $\Gamma\in L$, but
the order of integrabiity of the system depends
on the choice of the dynamical field, because:

a)  $(M,L,\Gamma)$ is Lie integrable  of order 2 (the minimum possible value) for
$\Gamma=X_i, i=1,2,3$ or any combination of them.

b) $(M,L,\Gamma)$ is Lie integrable  of
order 3 (the maximum possible value according to the result of the corollary)  for $\Gamma=X_4$ (or any combination in which the
coefficient of $X_4$ does not vanish).

\section{Distributional integrability}

It is clear that the preceding  construction is  too rigid or restrictive, because there are simple examples which cannot be analysed in
the framework here considered.
For instance, the system in $\mathbb{R}^n$ with dynamical vector field
\begin{equation}
 \Gamma=f(x)\partial_1\Longleftrightarrow
\dot x^1= f(x),\quad  \dot x^2=0,\quad  \dots,\quad  \dot x^n=0,\label{fvf}
\end{equation}
can be easily solved by quadratures but
the vector fields of the natural choice
\begin{equation}
 L=\langle \Gamma,\partial_2,\dots,\partial_n\rangle\,,\label{Lfvf}
\end{equation}
 do not close on a real Lie algebra. Note however that if $f$ is a never vanishing function  the dynamical vector field $\Gamma$  is conformally equivalent to $\partial/\partial x^1$.
 Moreover, we pointed out before that we can also consider non-strict symmetries of the dynamics which means that the set of solutions is preserved but with a reparametrisation of the integral curves.
 This suggests to extend the framework by considering $C^\infty(M)$-modules of vector fields instead of  $\mathbb{R}$-linear spaces. The price to be paid is that we do not have Lie algebras
 of vector fields anymore. However the idea of the construction developed in our approach can be maintained as it was proved in \cite{CFGR15}. We quickly sketch the generalisation developed in \cite{CFGR15}
 and refer the interested reader  to such paper.

First, for any subset $S\subset{\mathfrak{X}}(M)$, let $\mathcal{D}_S$ denote  the
$C^\infty(M)$-module generated by $S$:
$$
 \mathcal{D}_S=\left\{\sum_i f^i X_i\in {\mathfrak{X}}(M) \mid f^i\in C^\infty(M),\, X_i\in S\right\}.
$$
As $\mathcal{D}_S$ is the module of vector fields in the corresponding generalised distribution, we will also refer to $\mathcal{D}_S$ as to a distribution.

We say that a real vector space, $V\subset{\mathfrak{X}}(M)$, is \emph{regular}
if  $V$ is isomorphic to its restriction,  $V_p\subset T_pM$, at any point $p\in M$, and \emph{completely regular} if it is regular and  $V_p=T_pM$.

One basic definition is the following:

\begin{definition}
Given a completely regular vector space, $V\subset{\mathfrak{X}}(M)$, and a subset, $S\subset\mathfrak{X}(M)$, we shall call \emph{core} of $S$ in $V$,
denoted by $\core{S}$, the  {\it smallest} subspace of $V$ such that $S\subset \mathcal{D}_{\core{S}}$.
\end{definition}

One can prove that such a smallest subspace does exist: any subset
of ${\mathfrak{X}}(M)$ has a core.
This concept  of core of a generalised distribution
is  essential to extend the strategy for integration by quadratures
 from the Lie algebra setting to that of the $C^{\infty}(M)$-module
case.

First, in full analogy to the Lie integrability property, we introduce the
concept of {\it distributional integrability}.


Let be $V\in {\mathfrak{X}}(M)$ be a completely regular vector space  and  $\Gamma\in V$  a dynamical vector field.
We introduce the following sequence: $V_{\Gamma,0}=V$ and
$$
 V_{\Gamma,m}=\langle \Gamma\rangle + \core{[V_{\Gamma,m-1},V_{\Gamma,m-1}]} \,.
$$
We always have $V_{\Gamma,m}\subset V_{\Gamma,m-1}$.

The sequence  $V_{\Gamma,k}$ coincides with previously introduced
$L_{\Gamma,k}$ when $V=L$ closes a real Lie algebra.  In fact, one easily sees that in this case
$\core{[V_{\Gamma,m-1},V_{\Gamma,m-1}]}=[V_{\Gamma,m-1},V_{\Gamma,m-1}]$. It will play  a similar role in
 the more general case we are considering

\begin{definition}
{We say that  $(M,V,\Gamma)$ is \emph{distributionally  integrable of order $k+1$}} if $V_{\Gamma,k}$ is the first Abelian
(with respect to the commutator of vector fields) linear subspace in the decreasing sequence
$$V_{\Gamma,0}\supset V_{\Gamma,1}\supset V_{\Gamma,2}\supset\dots\ \,.$$
\end{definition}

We can now  state the main result of this section \cite{CFGR15}.

\begin{theorem}
If $(M,V,\Gamma)$ is distributionally integrable of order $r$,
then the  vector field $\Gamma$, can be integrated by $r$ quadratures.
\end{theorem}

Two examples were used in \cite{CFGR15} to illustrate the theory.
The first example is mentioned at the beginning of this section and explicitly given by (\ref{fvf}),
and then $V$ is given by the right hand side of (\ref{Lfvf}), i.e.
$V=\langle \Gamma,\partial_2,\dots,\partial_n\rangle$.
Then, we immediately see that
$[\Gamma,\partial_i]\in  \mathcal{D}_{\langle \Gamma\rangle}$ for any $i$,
and therefore $V_1=\langle \Gamma\rangle$,
so the system of equations is solved with two quadratures.

As a second example   (it requires  $n$ quadratures),
we can consider
$$
\Gamma=f(x)\big(
\partial_1+ g^2(x^1)\partial_{2}+\dots+
g^{n-1}(x^1,\dots,x^{n-2})\partial_{{n-1}}+g^{n}(x^1,\dots,x^{n-1})\partial_{n} \big),
$$
with $f(x)\not=0$ everywhere and $V=\langle \Gamma,\partial_{2},\dots,\partial_{n}\rangle$.
In this case,
$V_{\Gamma,1}=\langle \Gamma,\partial_{3},\dots,\partial_{n}\rangle$,
$V_{\Gamma,2}=\langle \Gamma,\partial_{4},\dots,\partial_{n}\rangle$,
and finally $V_{\Gamma,n-1}=\langle \Gamma\rangle.$
This shows that the system is distributionally integrable
and requires $n$ quadratures for its solution.

Remark  the appearance of a function $f$ multiplying the dynamical vector field in the previous examples.
This is, actually, the general situation as it was proved in \cite{CFGR15}.

\begin{proposition}
Suppose that  $(M,V,\Gamma)$, with
$V=\langle \Gamma,X_2,\dots,X_n\rangle$,
is distributionally integrable of order $r$.
Then, for any nowhere-vanishing $f\in C^\infty(M)$,
 the system $(M,V',f\Gamma)$ with
$V'=\langle f\Gamma,X_2,\dots,X_n\rangle$
is distributionally integrable of order $|r'- r|\leq 1$.
\end{proposition}

The conformally related vector fields $\Gamma$ and $f\,\Gamma$ have the same constants of motion, and therefore
the unparametrised orbits of both vector fields coincide \cite{CaIbLa88,Marle12}. In other words, as the integral curves of both are related by a time-reparametrisation,
 we can interpret the change of dynamical vector field from $\Gamma$ to $f\,\Gamma$  as a local, position dependent, redefinition of time.
Consequently,  our formalism allows for such arbitrary changes of time,
a property that it is  not true neither  in the Arnold--Liouville nor  in the the standard Lie theory of integration by quadratures.

\subsection*{Acknowledgments}
Financial support of the research projects \ MTM2015-64166-C2-1-
P, \ FPA-2015-65745-P (MINECO, Madrid), \ DGA-E24/1, E24/2 (DGA, Zaragoza) and
DEC-2012/06/A/ST1/00256 (Polish National Science Centre grant) is acknowledged.


\begin{thebibliography}{CFGR15}


\bibitem[AM80]{AbrahamMar} R. Abraham and J.E. Marsden,
{\sl Foundations of Mechanics}, Foundations of
Mechanics,  Second Edition, 1980.
\bibitem[AM88]{AbrahamMarRat} R. Abraham, J.E. Marsden and T. Ratiu,
{\sl  Manifolds, tensor analysis, and applications},
2nd edition, Applied Mathematical Sciences {\bf 75}. Springer-Verlag, New York, 1988.
\bibitem[A]{arnold} V.I. Arnold, {\sl Mathematical methods of classical mechanics}. Graduate Texts in Mathematics {\bf 60}, second edition, Springer,  1989.
\bibitem[AKN]{arnoldkozlov} V.I. Arnold, V.V. Kozlov and A.I. Neishtadt,  {\sl Mathematical aspects of classical and celestial mechanics}. Encyclopedia of Math. Sciences {\bf  3}, Springer, Berlin, 1989.
\bibitem[CCR]{CCR13}R.  Campoamor-Stursberg, J.F. Cari\~nena and M.F. Ra\~nada,
\emph{Higher-order superintegrability of a Holt related potential},
J.  Phys. A:Math. Theor. \textbf{46} (2013), 435202.
 \bibitem[CFGR]{CFGR15} J.F. Cari\~nena, M. Falceto, J. Grabowski and M.F. Ra\~nada,
 \emph{generalised Lie approach to integrability by quadratures},
 J.  Phys. A:Math. Theor. \textbf{48} (2015),  215206.
 \bibitem[CGL1]{CGdL12} J.F. Cari\~nena, J. Grabowski and  J. de Lucas, \emph{Superposition rules for higher-order systems and their applications},
J.  Phys. A: Math. Theor. {\bf 45} (2012), 185202.
\bibitem[CGM1]{CGM00} J.F. Cari\~nena, J. Grabowski and  G. Marmo, \textsl{Lie--Scheffers systems: A geometric approach},
Bibliopolis,  Napoli,  2000.
\bibitem[CGM2]{CGM01} J.F. Cari\~nena, J. Grabowski and  G. Marmo,
\emph{Some physical applications of  systems  of differential equations
admitting a superposition rule}, Rep. Math. Phys. {\bf 48} (2001), 47--58.
\bibitem[CGM3]{CGM07} J.F. Cari\~nena, J. Grabowski and  G. Marmo, \emph{Superposition rules, Lie theorem and partial differential equations},
Rep. Math. Phys. {\bf 60} (2007), 237--258.
 \bibitem[CGR]{CGR01}J.F. Cari\~nena, J. Grabowski and  A. Ramos, \emph{Reduction of time--dependent systems admitting a
 superposition principle}, Acta Applicandae Mathematicae {\bf 66} (2001), 67--87.
\bibitem[CI]{CI88} J.F. Cari\~nena and L.A. Ibort,
\emph{Noncanonical groups of transformations, anomalies and   cohomology},
 J. Math. Phys. \textbf{29} (1988), 541--545.
\bibitem[CIL]{CaIbLa88} J.F. Cari\~nena, L.A. Ibort and E.A. Lacomba, {\it Time scaling as an infinitesimal canonical transformation in Celestial Mechanics}, Celest. Mechanics \textbf{42} (1988), 201--213.
 \bibitem[CIMM]{CIMM15} J.F. Cari\~ nena,  A. Ibort, G. Marmo and  G. Morandi, {\sl Geometry from Dynamics, Classical and Quantum},
  Springer, 2015.
\bibitem[CL1]{CdLa} J.F. Cari\~ nena and  J. de Lucas, \emph{Integrability of Lie systems through Riccati equations},
Int.  J. Nonlinear Mathematical Physics,   {\bf 18} (2011), 29--54.
\bibitem[CL2]{CdLb} J.F. Cari\~ nena and  J. de Lucas, \emph{Lie systems: theory, generalisations, and applications}
Dissertationes Mathematicae {\bf 479},  Institute of Mathematics, Polish Academy of Sciences, Warszawa, 2011.
\bibitem[CLR]{CdLR} J.F. Cari\~ nena, J. de Lucas and   A. Ramos, \emph{A geometric approach to integrability conditions for  Riccati
equations}, Electronic J. Diff. Equations  {\bf  122} (2007),  1--14.

\bibitem[COW]{COW93} J.F.  Cari\~nena, M.A. del Olmo and P. Winternitz,  \emph{On the relation between weak and strong invariance of
differential equations},
 Lett. Math. Phys. \textbf{29} (1993), 151--163.
\bibitem[CR]{CR99} J.F. Cari\~ nena and A. Ramos, \emph{Integrability of Riccati equation
from a group theoretical viewpoint}, Int. J. Mod. Phys. {\bf A 14}  (1999), 1935--1951.

\bibitem[CE]{CE48} C. Chevalley and S. Eilenberg,
\emph{Cohomology theory of Lie groups and Lie algebras},
Trans. Amer. Math. Soc. \textbf{63}  (1948), 85--124.
\bibitem[CP]{Crampinbook} M. Crampin and F.A.E.  Pirani,
{\sl Applicable Differential Geometry}. Cambridge University Press, 1986.
\bibitem[G]{G90} J. Grabowski,
\emph{Remarks on nilpotent Lie algebras of vector fields},
J. Reine Angew. Math. \textbf{406} (1990), 1--4.
\bibitem[H]{H82}  C.R. Holt,
\emph{Construction of new integrable Hamiltonians in two degrees of freedom},
J. Math. Phys.  \textbf{ 23} (1982),   1037--1046.
\bibitem[MK]{MK88} M. Kawski,
\emph{Nilpotent Lie algebras of vector fields},
J. Reine Angew. Math. \textbf{388} (1988),  1--17.
\bibitem[VVK1]{K83} V.V. Kozlov, \emph{Integrability and nonintegrability in Hamiltonian mechanics},
Russian Math. Surveys \textbf{38} (1983), 1--76.
\bibitem[VVK2]{K05} V.V. Kozlov,  \emph{Remarks on a Lie theorem on the integrability of differential equations in closed form},
Diff. Eqns {\bf 41}(4)  (2005), 588--590.
\bibitem[L]{JLiou53} J. Liouville,   \emph{Note sur l'int\'egration des \'equations diff\'erentielles de la dynamique}, pr\'esent\'ee au Bureau des longitudes le 29 juin 1853, Journal de Math\'ematiques pures et appliqu\'ees {\bf 20}  (1855),  137--138.
\bibitem[M]{Marle12} C-M. Marle, \emph{A property of conformally Hamiltonian vector fields; application to the Kepler problem},
J. Geom. Mech. \textbf{4} (2012), 181--206.
\bibitem[MF] {MF78}A. Mishchenko and A. Fomenko,
\emph{generalised Liouville method of integration of Hamiltonian systems}, Funct. Anal. Appl.   \textbf{12} (1978), 113--121.
\bibitem[MCSL]{MCSL14} R. Mohanasubha, V. K. Chandrasekar, M. Senthilvelan and M. Lakshmanan,
\emph{Interplay of symmetries, null forms, Darboux polynomials, integrating factors and Jacobi multipliers in integrable second-order differential equations},
Proc. R. Soc. A   \textbf{470} (2014), 20130656.
\bibitem[O]{olver} P.J. Olver, \textsl{Applications of Lie Groups to Differential Equations}, Berlin: Springer, 1986.
\bibitem[PW]{PW11} S. Post  and P. Winternitz,
\emph{A nonseparable quantum superintegrable system in 2D real Euclidean space},
J. Phys. A: Math. Theor. \textbf{44}  (2011), 162001.

\end{thebibliography}
\end{document}